# Calibration-based Minimalistic Multi-Exposure Digital Sensor Camera Robust Linear High Dynamic Range Enhancement Technique Demonstration

NABEEL A. RIZA AND *NAZIM ASHRAF

*School of Engineering, University College Cork, Cork, Ireland*
*Corresponding author: n.riza@ucc.ie*

**Demonstrated for a digital image sensor-based camera is a calibration target optimized method for finding the Camera Response Function (CRF). The proposed method uses localized known target zone pixel outputs spatial averaging and histogram analysis for saturated pixel detection. Using the proposed CRF generation method with a 87 dB High Dynamic Range (HDR) silicon CMOS image sensor camera viewing a 90 dB HDR calibration target, experimentally produced is a non-linear CRF with a limited 40 dB linear CRF zone. Next, a 78 dB test target is deployed to test the camera with this measured CRF and its restricted 40 dB zone. By engaging the proposed minimal exposures, weighting free, multi-exposure imaging method with 2 images, demonstrated is a highly robust recovery of the test target. In addition, the 78 dB test target recovery with 16 individual DR value patches stays robust over a factor of 20 change in test target illumination lighting. In comparison, a non-robust test target image recovery is produced by 5 leading prior-art multi-exposure HDR recovery algorithms using 16 images having 16 different exposure times, with each consecutive image having a sensor dwell time increasing by a factor of 2. Further validation of the proposed HDR image recovery method is provided using two additional experiments, the first using a 78 dB calibrated target combined with a natural indoor scene to form a hybrid design target and a second experiment using an uncalibrated indoor natural scene. The proposed technique applies to all digital image sensor-based cameras having exposure time and illumination controls. In addition, the proposed methods apply to various sensor technologies, spectral bands, and imaging applications.**

## 1. Introduction

Linear HDR imaging [1-2] is critical for deciphering low contrast targets within HDR scenes, including enhancing multispectral imaging reconstructions. Furthermore, accurate and reliable quantitative image capture is important for mission critical applications where incorrect image data can lead to inaccurate image recovery and hence catastrophic system failure. One such critical application is medical imaging where one deploys silicon multi-pixel digital image sensors. In general, cameras designed using classic photo-electron storing digital image sensors such as CMOS/CCD/FPA image sensors via their device and circuit physics inherently produce a non-linear CRF over the HDR (e.g., 90 dB), in particular in the low light and bright light regions [3]. Recent experiments indeed show such a non-linear CRF prevents a deployed commercial HDR CMOS camera from registering differential output signals with adequate Signal-to-Noise Ratio (SNR) for capture of low contrast targets within a calibrated HDR scene [4].

A classic multi-exposure approach for Dynamic Range (DR) extension was proposed in 1962 [5]. This approach was initially developed in the late 80's [6-8] and early 90's [9-10] to enhance digital sensor CCD-based camera DR. This fundamental multi-exposure DR extension approach used multiple un-calibrated synthesized or real images of different sensor exposure time values that were engaged with an algorithm to first generate the CRF. Next deployed was the full range CRF with real capture multi-exposure images to generate an HDR image that was otherwise not possible using a single exposure limited DR image. Since the mid-90's to date, various types of leading multi-exposure algorithms for DR extension have been proposed that calculate the CRF from acquired image data and use different weighting schemes to find the final scaled pixel irradiances of the HDR image. These weighting schemes vary from using the slope of the CRF at its specific irradiance level [11] to using a hat shaped weighting function [12] to using a weighting function that is dependent on the SNR and the CRF's derivative [13]. In addition, a weighting scheme deployed a Gaussian function coupled with the sensor exposure time [14] and another very recent weighting approach used a rank minimization algorithm [15] deployed with synthesized multi-exposure Low Dynamic Range (LDR) image data to recover the HDR image. Furthermore, others have used a recursive filter weight map [16] and a weight guided imager filter for Gaussian pyramid weight smoothing [17]. Researchers have also proposed an image processing method that avoids the required physical CRF computation and instead guides multi-exposure image fusion by using contrast and saturation quality measures for HDR generation [18]. Apart from image processing algorithms work, the use of improved pixel electronics hardware design has also been attempted to improve CMOS image sensor linearity [19]. Nevertheless, achieving full DR camera linearity as well as robust CRF generation remains a challenge to enable robust HDR image recovery using LDR multi-exposure camera operations.

It has recently been shown by Riza and Ashraf in Ref.20 that by using known calibrated HDR targets to experimentally determine the best estimate of the true CRF, some leading prior-art multi-exposure algorithms produce non-robust HDR images [20]. Part of this non-robustness comes from the inaccuracy of the deployed CRF as well as the fact that these algorithms engage the full CRF range data that fails to maintain linearity required for ideal multi-exposure HDR image generation operations [11-15]. Ref.20 also introduced a calibration empowered HDR image generation method that restricts data processing operations to an optimal smaller DR linear CRF range and engages ideal unweighted multi-exposure processing with required camera HDR design parameter dependent minimal images to produce improved robustness captured HDR image data. This paper describes further limitations of the ref.20 deployed CMOS sensor technology and demonstrates additional critical optimizations required for an experimentally calibrated CRF generation, leading to an improved accuracy and reliability of the observed HDR image when compared to HDR image generation using five leading prior-art multi-exposure image processing algorithms [11-15].

The paper starts by describing the proposed optimized CRF generation technique for a HDR CMOS image sensor-based camera. Using a 90 dB calibration target, an experiment is conducted to generate the required CRF. This CRF is next used for multi-exposure image processing-based HDR image generation via the proposed minimal exposures high linearity method and compared with experimental imaging results obtained using 5 leading prior-art HDR enhancement multi-exposure algorithms. The paper concludes with a summary of the proposed methods and its experimental results that showcase the image recovery robustness advantages.

## 2. Proposed Digital Sensor Camera CRF Generation and Multi-Exposure Imaging Techniques

A digital image sensor is a hard-wired multi-pixel light sensing optoelectronic device with individual custom semiconductor material (e.g., silicon, indium gallium arsenide, mercury cadmium telluride, lead selenide, indium antimonide, etc) light absorbing photo-cells arranged in a spatial grid with cells having specific incident photo-charge collection and timing control circuitry. For visible light and an HDR scenario, CMOS-based silicon sensors are a dominant technology and hence the focus of ref.20 and this follow-on paper. There are number of different hardware methods to realize a CMOS sensor with HDR performance [21-24]. These designs inherently have a non-linear input-output relationship between incident light intensity on a pixel (i.e., photo-cell) and the pixel output voltage read by the sensor readout circuitry. This optoelectronic light-to- electrical input/output transducer relationship is at best a pseudo-linear input-output function that using ideal camera optics transfers linearly to the camera (i.e., sensor plus lens optics) in-out performance designation called the CRF.

As mentioned in the introduction, digital CMOS sensors at their extreme maximum and minimum incidence light level zones become non-linear, thus restricting the full HDR input light range for full range linear DR capture per single exposure photo-shot of the camera. Hence multi-exposure multi-shot techniques and algorithms were proposed in the mid 1980's [6-8] and now are commonly deployed to extend the DR of an otherwise limited linear DR CMOS sensor-based camera. In such cases as proposed in ref.20, not only is a "true" calibrated experimental CRF required for ideal multi-exposure image processing, one must also restrict the obtained CRF usage to only the best continuous linear region. In addition, ref.20 also proposes to calibrate the camera over tested overall illumination and time factor exposure ranges that ensures that the camera maintains linearity of the input-output mapping required for ideal linear DR extension via multi-exposure image processing for an otherwise pseudo-linear response camera.

The proposed calibration process starts with the choice of the selected digital sensor camera to be deployed using multi-exposure image processing for a specific linear HDR imaging application for low contrast detection with a desired design linear HDR value in dB called $HDR_D$. In other words one requires multi-exposure processing as the chosen camera has an instantaneous single-shot dynamic range called $HDR_I$ that is less than the desired full linear $HDR_D$ value. For example the camera manufacturer specified $HDR_I$ = 95 dB while the desired $HDR_D$ =135 dB. Furthermore, the specified $HDR_I$ is not necessarily a linear DR and therefore requires the proposed calibration procedure to identify the camera's experimental linear DR called $LDR_E$ in dB.

To implement this test, one first needs to design a calibrated multi-section target with a maximum DR ≥ $HDR_I$ . Because low contrast detection is required, near 6 dB differential DR detection (or 2:1 difference between irradiance values) between imaged pixels should be achieved [20]. Hence the designed camera calibration test target must have differential target spatial zones that vary in near 6 dB steps over the full $HDR_I$. The number of designed different DR zones on the calibration target is ≥ $HDR_I$ / 6 with the brightest light zone on target representing the 0 dB DR marker and the darkest light zone on target representing the $HDR_I$ marker. The choice of the spatial averaging zone size of an observed test target for calibration depends on the optical parameters of the camera such as field-of-view, demagnification factor, pixel size, pixel count as well as the required number of different DR zones on the calibration test target. To get an adequate spatial averaging over a specific DR value target zone, a minimum of 0.1% of the total pixel count in the digital sensor, e.g., 1000 pixels in a 1 million pixels sensor can be deployed which in-turn sets the individual zone area size. The inter-zone spaces between the calibrated DR value zones should be the black or no light emission/reflectance regions of adequate size relative to DR zones to minimize inter-zone optical crosstalk between the imaged test zones on the digital sensor.

The CRF calibration generation process starts by using a bright and high uniformity light source illuminating the designed calibration target with the test camera used to image the in-focus calibration target. Today's digital sensors have many (e.g., million or more) tiny (e.g., < 10 micron square) pixels (photo-cells) that during bright light exposure easily saturate and often display a saturated pixel triggering anomaly for a small fraction of pixels in the sensor. The spatial averaging process of the pixel voltages $v_p$ values over a test zone also counters this anomaly and is used to measure the sensor output voltage response to a given input light level at the specific DR value test zone.

To set the exposure time of the sensor for the single exposure (or single shot) single image calibration process, the exposure time should be set to the value such that the averaged $v_p$ value for the 0 dB test zone (i.e., brightest zone) is closest to the camera $v_p$ maximum value. For example, if the camera sensor is specified as a A-bit output sensor, the maximum average $v_p$ output value one tries to measure with exposure control is $v_p(avg)= 2^A -1$. Furthermore, a pixel output $v_p$ histogram analysis for the target averaged test region for each designed DR value zone is conducted to check which brightest zone in the calibration target has no individual pixel saturations as this DR value marks when the camera is no longer in the non-robust non-linear CRF regime.

Given the calibration target has same size test target zones for specific design DR values illuminated by a uniform illumination (e.g., < 5% variation), the camera provided single shot image gives $v_p(avg)$ values per zone that then allows one to plot the CRF, i.e., $v_p(avg)$ versus scaled irradiance value curve. Note that the DR values can be used to compute a scaled irradiance value given the 0 dB test DR zone represents the brightest light in the calibration image matching the $v_p(avg)= 2^A-1$ max value of the sensor. Depending on the linearity level (i.e., percentage line slope variation tolerance) desired, analysis of the CRF curve can provide a measure of the tested camera $LDR_E$ that in-turn determines the limits to the use for the camera provided $v_p$ values between a $v_{max}$ value and a $v_{min}$ value where the camera operates in the desired linear regime.

With the experimental CRF measured giving the $LDR_E$ value, the camera is ready for execution of the proposed minimal images weighting-free multi-exposure linear HDR extension method. Specifically, for camera operations providing a $HDR_D$ multi-image exposure processing performance, one must satisfy the condition N x $LDR_E \geq HDR_D$, where N is the minimum number of different multi-exposure images required. Here, the shortest exposure $T_1$ value is chosen such that the brightest zone of the observed scene (e.g., the 0 dB target patch zone of a deployed test scene) produces a spatially averaged $v_p(avg)$ value called $v_B$ that satisfies $v_B \leq v_{max}$ to ensure that one is operating in the linear CRF regime. Note that current digital sensors exhibit intrinsically non-linear CRF at the brighter light levels hence the proposed $v_B \leq v_{max}$ is required. Typically, one would expect the brightest zone in the observed scene to be covering many sensor pixels and hence averaging of these brightest zone $v_p$ values can produce a robust measurement of $v_p(avg)$ to meet the condition $v_p(avg)=v_B<v_{max}$.

The next longer sensor exposure time $T_2$ used to get the second image using the full $LDR_E$ from the camera is obtained with $T_2 = P_2 T_1$ where the factor $P_2$ is computed from $20 \log P_2 = LDR_E$. This second exposure gives the second image used for the multi-exposure image processing execution without the use of any image weighting functions (unlike prior-art) to combine image data as the proposed multi-image acquisition process has maintained camera linearity throughout the image capture processes. If a third exposure $T_3$ using the camera full $LDR_E$ is required for continuing to achieve the $LDR_D$ value, then $T_3= P_3 T_2$ where the factor $P_3 = P_2$. This process can be continued for a total of N acquired images with N different time exposures such that the $N^{th}$ exposure time $T_N=P_N T_{N-1}$ where the last and longest exposure time increase factor $P_N$ is computed from $20\log P_N = HDR_D – (N-1)LDR_E$. In general, $T_n$ is nth-image exposure times with n=1,2,. 3, .....,N.

It is also important to note that for each exposure time $T_n$, only pixel output $v_p$ values between $v_{max}$ and $v_{min}$ can be used for image generation as these pixel voltage data values fall in the robust linear CRF regime of the camera. In addition, simply increasing pixel exposure time beyond a certain limit for the sensor does not imply that the $v_p$ values will continue to increase in a linear fashion with increasing exposure times as at very low light levels, the digital sensor again enters a non-linear CRF regime where the proposed all-linear multi-exposure HDR extension technique will fail to produce the desired linear HDR extension.

## 3. Proposed Digital Sensor Camera CRF Generation Experiment

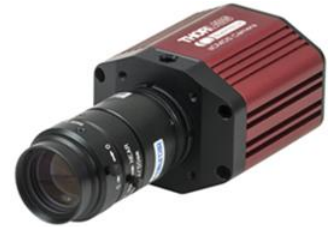

Fig. 1.  Deployed 87 dB HDR CMOS-sensor based camera.

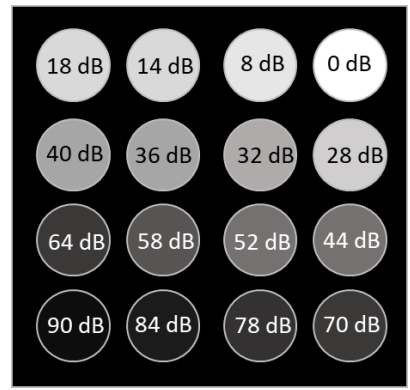

Fig. 2.  Design of the deployed 90 dB HDR calibration target for CRF generation.

To demonstrate the proposed CRF generation method, deployed is the Thorlabs monochrome CMOS sensor-based camera model Quantalux S2100-M with 5.04 μm pixel pitch, 2.1 Mpixels and up-to an $HDR_I$=87 dB rating with a A=16-bit (i.e., 0 to 65,535 levels) $v_p$ output. As shown in Fig. 1, the camera is fitted with a C-mount GMZ18108 lens that images the known DR test targets. The laboratory is airconditioned and maintains a steady cool room temperature via thermostat control, preventing large temperature fluctuations that could affect CMOS sensor behavior. The target is placed 168.7 cm from the sensor end of the camera.

Fig. 2 shows the design of the deployed 90 dB HDR calibration target that is made from a 16 = 4 x 4 grid of optical patches with different designed optical attenuation values. Note that the selected target maximum DR of 90 dB satisfies the > $HDR_I$ condition. In addition, the 16 target patch zones satisfy the low contrast detection calibration chart condition as 16 zones are used and 16>15 where (90 dB)/(6 dB) = 15 zones. Attenuation is implemented using circular shape Thorlabs Neutral Density (ND) filters with experimentally verified attenuation factors. The 16 patch target scene assembly is a 9.1 cm x 9.1 cm square area with an inter-patch distance of 1.45 cm with each target patch with a Thorlabs ND filter aperture size of 1.27 cm diameter. A black acetal sheet material is used between the patches that makes the test targets of the low glare type. The Fig. 2 patch with 0 dB label is the brightest patch while the 90 dB patch is the weakest light patch matched near the deployed camera DR rating of 87 dB.

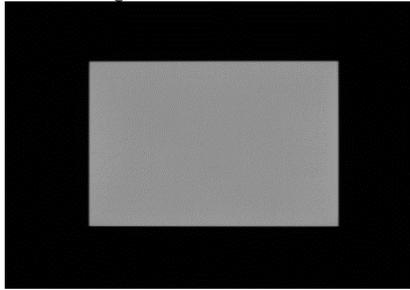

Fig. 3. CMOS Sensor-based camera captured LG3 uniform illumination zone (central white area) seen under non bright light conditions that as normally expected, generates no saturated pixel triggering in the captured scene.

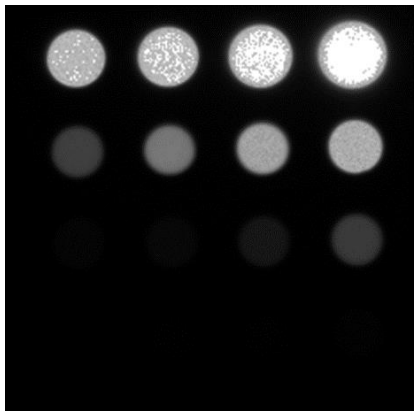

Fig. 4. Single shot image of the 90 dB calibration target used for CRF generation.

For accurate CRF generation, one must ensure that the exposure time of the camera is set to a value small enough to capture an image just under pixel saturation so a widest possible image DR can be recorded for a single shot. Under bright light 60 KLux illumination conditions using the Image Engineering (Germany) Model LG3 white light box illuminating the 16 patch 90 dB HDR calibration target, the Thorlabs 16-bit $v_p$ output signal CMOS camera for the brightest patch started to trigger pixel saturations (i.e., $v_p=2^{16}-1=65535$) for 0.93% of pixels in the 0 dB patch zone for a 0.296 ms exposure time. This 0 dB patch zone covering 14100 CMOS sensor pixels gives a computed mean $v_p$ called $v_p$(avg) of 39218. Ideally, this $v_p$(avg) should be 65535 as the CMOS sensor received a uniform patch of bright illumination, so all CMOS pixels in the patch should have produced a $v_p=65535$.

To counter this experimental anomaly, the raw acquired CRF $v_p$ data in Ref.20 was scaled by a factor of 1.67, as the $v_p$(avg)=39218 at the hint of pixel saturation is expected to be at the saturation value of 65535. In reality, such scaling ignores the highly nonlinear behavior of the deployed CMOS sensor for brighter light conditions where the onset of saturated pixel signal outputs had started. Because of this saturated pixel triggering behavior observed with brighter light conditions for the deployed CMOS sensor that maybe common place for CMOS & other digital sensor technologies, this paper proposes an improved CRF generation technique based on pixel $v_p$ output signal histogram analysis and spatial zone averaging given the use of a highly calibrated target. It is important to point out that under lower brightness light uniform illumination conditions, the deployed CMOS sensor-based camera does not exhibit saturated pixel triggering, such as shown in Fig.3 where no pixels saturate. In this case, the average pixel output reading spatially averaged over the captured CMOS sensor pixels white uniform LG3 illumination zone is 38537 and pixel data analysis gives a 96% homogeneity across the zone. Note that LG3 light box manufacture specifies a less than 5% variation (i.e., >95% homogeneity) of illumination levels across the entire illumination screen, indicating that the deployed CMOS camera meets the designed uniformity constraints. It is important to point out that the observed saturated pixel triggering effect is seen across many different regions of the CMOS sensor pixel grid. With a modified sensor exposure time or light level, these pixels no longer show the saturation triggering effect. In other words, such pixels in the deployed sensor are not product defects or artefacts that can simply be removed by image preprocessing.

Specifically, demonstrated is CRF generation using the Fig. 2 custom design HDR target similar to ref.20 target, but with a higher 90 dB DR and patch attenuation values in dB of 0, 8, 14, 18, 28, 32, 36, 40, 44, 52, 58, 64, 70, 78, 84, 90. The camera calibration exposure time $T_C$ is set to 3.703 ms as it gives the 0 dB brightest patch computed $v_p$(avg) of 64537.8 which is very near the 16-bit $v_p$ limit. Fig. 4 shows this single shot image of the 90 dB calibration target that is used for CRF generation. The $v_p$(avg) values for remaining patches are also computed and shown in Table 1. One can assume that the scaled input maximum irradiance value $I_s$ for the zero attenuation or 0 dB patch is $I_s=10^6$ for generating the CRF plot. For each patch having a known DR value, the equivalent $I_s$ value can be computed and is shown in Table 1. In addition, pixel $v_p$ value histogram analysis for all 16 patches is done that shows that the 28 dB patch is the first patch of the brighter patches to show no saturated pixels, i.e., no $v_p$ values of 65535.

Table 1. $v_p$ (avg) values measured for the 16 patches in the 90 dB calibration target.

| Design (dB) | Scaled Input Light Irradiance | 16-bit CMOS Output $v_p$(avg) |
|---|---|---|
| 0 | 1000000.0 | 64537.8 |
| 8 | 398107.2 | 59537.6 |
| 14 | 199526.2 | 52046.9 |
| 18 | 125892.5 | 47782.7 |
| 28 | 39810.7 | 42901.4 |
| 32 | 25118.9 | 37486.7 |
| 36 | 15848.9 | 26563.7 |
| 40 | 10000.0 | 18573.6 |
| 44 | 6309.6 | 13003.9 |
| 52 | 2511.9 | 5200.5 |
| 58 | 1258.9 | 2134.2 |
| 64 | 631.0 | 1438.2 |
| 70 | 316.2 | 804.4 |
| 78 | 125.9 | 486.5 |
| 84 | 63.1 | 389.6 |
| 90 | 31.6 | 285.5 |
| No Light (Black Zone) | 0.0 | 191.3 |

Fig. 5 shows histogram plots showing the number of CMOS sensor pixels having specific 16-bit scale individual pixel range $v_p$ signal outputs with a $v_p$ range of 1000. The top plot is for the 14 dB DR target patch showing 169 saturated pixels while bottom plot is for the 28 dB target patch that shows no saturated pixels. The Table 1 data is used to produce the Fig. 6 CRF plot that is engaged for multi-exposure imaging for linear DR extension. Note that given limits in computer-based quantization errors, an appropriate and sufficiently large should be used to allow accurate and robust slope computations between adjacent data points in plot. The absence of light inter-patch black region, i.e., for scaled $I_s$ = 0 value measures a $v_p(avg) = v_N$ of 191. Slope analysis between all adjacent data points shows that near continuous linear CRF behavior with an average slope value of 1.65 between the 32 dB patch with a $v_p(avg)$=37486 and the 84 dB patch with a $v_p(avg)$=389. This data in turn sets the maximum and minimum $v_p$ limits for multi-exposure data image processing that ensures a linear CRF mapping is maintained. Specifically, one gets $v_{max}$ = 37486 and $v_{min}$ = 389. The lowest SNR occurs for the $v_p(avg) = v_{min}$ value with an SNR= $v_{min}/v_N$ = 389/191= 2.

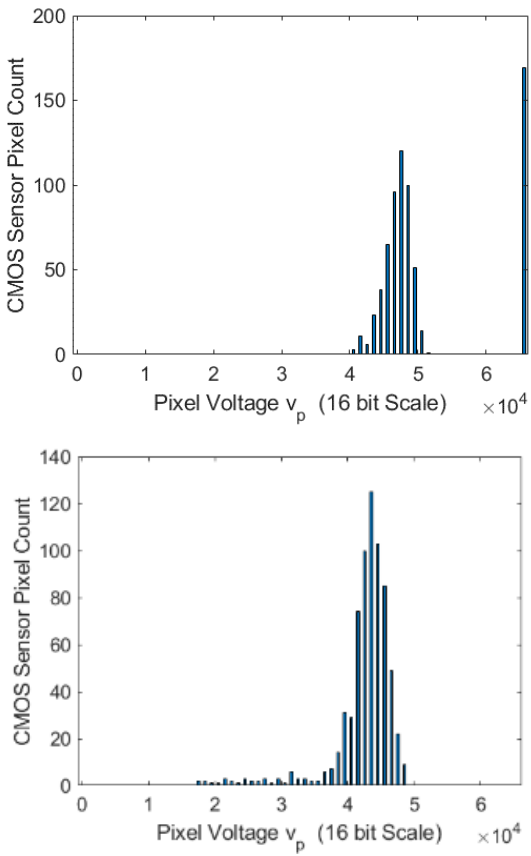

Fig. 5. Histogram plots showing the number of CMOS sensor pixels having specific 16-bit scale individual pixel range $v_p$ signal outputs. Top plot: 14 dB DR target patch with 169 saturated pixels. Bottom plot: 28 dB target patch with no saturated pixels.

Histogram data analysis of individual $v_p$ values in the single shot image also confirms that the near continuous slope value linear CRF behavior occurs for patches with DR> 28 dB, thus avoiding any nonlinear effects due to the saturated pixel triggering anomaly observed in the current digital CMOS sensor. Note that for any individual pixel $v_p$ > 64537, an $I_s$=$10^6$ is assigned. Similarly, for any individual pixel $v_p$<191, an $I_s$=0 is assigned. Thus, the measured experimental camera linear dynamic range for this camera is $LDR_E$ = 20log($v_{max}$ / $v_{min}$) = 20log(37486/389) = 39.66 dB, although the camera specifications indicate a DR up-to 87 dB. Also note that CRF generation robustness will further improve if the calibration image is taken multiple times and then averaged to get the final image deployed for CRF generation.

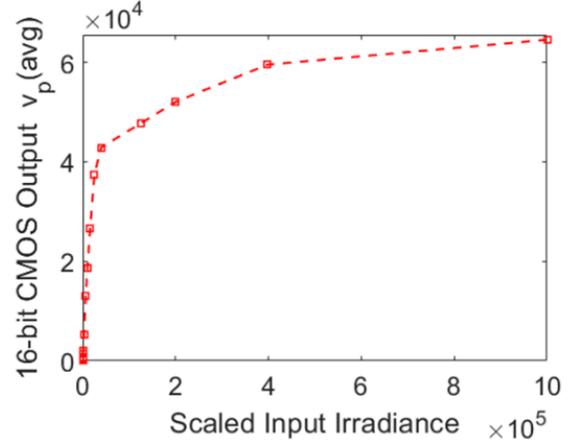

Fig. 6. Experimental CRF plot for the tested CMOS sensor-based camera.

## 4. Proposed Minimalistic Multi-Exposure Linear HDR Imaging Technique Experiment

Given the present camera has a near 40 dB experimental $LDR_E$, using a minimum of N=2 multi-exposure images implies that a $HDR_D$ = 80 dB designed linear HDR target recovery is possible as N x $LDR_E$ ≥ $HDR_D$. Hence as a first fundamental step in experimental verification of the proposed CRF calibration and multi-exposure unweighted image processing linear HDR extension methods, a designed test target of 78 dB DR is deployed so the minimum N=2 images can be used for linear DR extension. $P_2$=100 factor between exposure times for the two images where $T_N$= $P_2$ $T_1$. Recall that using N=2, 20log $P_2$ = $HDR_D$ – $LDR_E$ and the shorter exposure $T_1$ value is chosen such that the brightest known test target produces a spatially averaged $v_p(avg)$ value called $v_B$ that satisfies $v_B$ ≤ $v_{max}$ = 37486. Similarly, the final $n^{th}$ image (n=N, an integer) is taken with the longest $T_N$ exposure time with a $v_p(avg)$ value called $v_W$ that meets the condition $v_W$ ≥ $v_{min}$=389. To generate the final linear HDR recovered image, the experimental CRF in Fig. 6 along with the time factor $P_2$ value of 100 and the individual pixel $v_p$ values of the N=2 acquired images are used to produce the individual CMOS pixel scaled irradiance values of the final linear HDR composite image.

Given that the concluded camera system design calculations using the Fig. 6 experimental CRF indicates a 80 dB linear HDR image recovery potential, the Fig. 7 test target design with a maximum 78 dB DR and again using 16 patch zones with low contrast step DR values is deployed to test the designed minimal multi-exposure unweighted image linear DR extension method with the newly calibrated CMOS sensor-based camera in the laboratory. But before demonstrating the proposed minimal images weighting-free multi-exposure technique for linear HDR recovery, it was relevant to use the measured CRF with other well-known multi-exposure HDR imaging methods. Unlike the proposed method, these prior methods do not deploy specific restrictions on exposure times and $v_p$ values and rely on acquiring several images following an exposure scheme, like a factor of 2 increase between consecutive exposures. These images are acquired such that both the weakest light pixel values are captured above the camera noise floor using the longest exposure setting and the brightest pixel values in the scene are captured under saturation using the shortest exposure time. Images between the shortest and longest exposure times have pixel values of the scene between the extreme values.

**Table 2. Test 78 DB HDR image recovery using 16 multiple exposure images using 5 prior-art leading algorithms as well as the proposed method with the non-optimal image count of 16.**

| Design (dB) | Proposed [20] | UC Berkeley [12] | Notre Dame [14] | Sony/Columbia [13] | MIT [11] | KAIST [15] |
|---|---|---|---|---|---|---|
| 0 | 0 | 0 | 0 | 0 | 0 | 0 |
| 8 | 8.8 | 8.8 | 0.6 | 13.2 | 11.9 | 1.6 |
| 14 | 14.7 | 20.4 | 5.3 | 23.8 | 19.4 | 3 |
| 20 | 22 | 35 | 16.2 | 37.2 | 27.6 | 4.7 |
| 26 | 26.3 | 40.1 | 21.3 | 42.5 | 32.2 | 6.5 |
| 32 | 34 | 47.7 | 26 | 51.5 | 40 | 10.4 |
| 36 | 37.9 | 51.8 | 28.4 | 56.3 | 44.6 | 13 |
| 40 | 41.7 | 54.8 | 29.5 | 60 | 48.1 | 14.5 |
| 44 | 44.9 | 57.3 | 30.5 | 63.4 | 51.6 | 15.4 |
| 50 | 55 | 63.6 | 32.2 | 71.3 | 61.2 | 20.4 |
| 56 | 58.1 | 65.6 | 32.7 | 73.7 | 64.4 | 21.8 |
| 60 | 63.1 | 68.8 | 33.3 | 77.8 | 69.9 | 23.7 |
| 64 | 66.5 | 71.2 | 34 | 81 | 73.8 | 24.7 |
| 68 | 69 | 73 | 34.5 | 83 | 76.9 | 25.2 |
| 74 | 73.1 | 75.8 | 35.3 | 86.1 | 81.6 | 25.5 |
| 78 | 75.8 | 77.9 | 36 | 88.4 | 84.7 | 25.7 |

For example, a large number, namely, 16 images were captured by the present camera with a time factor scaling of 2 using the following exposure settings in ms of 0.029, 0.059, 0.118, 0.237, 0.474, 0.948, 1.896, 3.792, 7.585, 15.17, 30.340, 60.681, 121.362, 242.725, 485.451, 970.903. Table 2 provides the recovered test image DR values using 5 leading multi-exposure algorithms [11-15] and compares it with the proposed method using the un-optimal large count 16 images versus the optimal minimal 2 images. In the Mann and Picard approach [11], each pixel measured scaled irradiance is weighted by the slope of the CRF at its specific irradiance level and final scaled pixel irradiance is the average of the processed N images. Debevec and Malik [12] use the following hat function to weigh each pixel's measured scaled irradiance in an acquired image:

$$w(z) = \begin{cases} z - Z_{min} \; for \; z \leq \frac{1}{2}(Z_{min} + Z_{max}) \\ Z_{max} - z \; for \; z > \frac{1}{2}(Z_{min} + Z_{max}) \end{cases} \quad (1)$$

where $Z_{min}$, and $Z_{max}$ are lowest and highest possible pixel values, respectively. For the deployed CMOS sensor, these values are 0 and 65535 for $Z_{min}$, and $Z_{max}$, respectively. This weighting is applied in the logarithmic domain and the final scaled pixel irradiance is the inverse logarithm of the weighted average of these scaled irradiances. This weighting scheme is designed to give higher weightage to mid-range pixel values and less importance to pixel values at the ends of the sensor output range. Mitsunaga and Nayar [13] use the SNR as the weighting for the pixel's measured scaled irradiance and deploy the ratio of the CRF's derivative, i.e., CRF' at the specified irradiance level. The final scaled pixel irradiance is the average of these scaled irradiances. In summary, pixel values with higher SNR get higher weightage and vice versa. Robertson, Borman, & Stevenson [14] use a weighting scheme similar to Debevec and Malik [12] coupled with the exposure time. Instead of a hat function, they use a Gaussian-like function given as:

$$w(z) = exp\left[-W \cdot \frac{(z - Z_{mid})^2}{(Z_{mid})^2}\right]$$

**Table 3. Recovered HDR 78 DB target patch data using the proposed multi-exposure HDR recovery technique.**

| Design (dB) | Measured by proposed HDR Recovery Method |
|---|---|
| 0 | 0 |
| 8 | 10.1 |
| 14 | 15.7 |
| 20 | 21.9 |
| 26 | 26.5 |
| 32 | 35.2 |
| 36 | 40.2 |
| 40 | 41.8 |
| 44 | 43.7 |
| 50 | 53.5 |
| 56 | 56.5 |
| 60 | 61.1 |
| 64 | 64.1 |
| 68 | 68 |
| 74 | 73.4 |
| 78 | 77.4 |

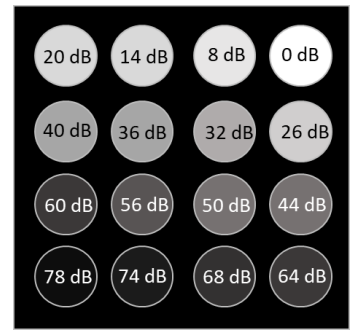

Fig. 7. 78 dB test target design for linear HDR image generation using deployed CMOS Camera.

The Gaussian function is scaled and shifted so that w(0) = w($Z_{max}$) = 0 and w($Z_{mid}$) = 1 where $Z_{max}$, and $Z_{mid}$ are highest and the middle pixel value, respectively. For the camera used in this paper, the values were 65535 and 32768 for the $Z_{max}$ and $Z_{mid}$, respectively. W is a numerical value that represents the confidence in the reliability of pixel observations. The final scaled pixel irradiance is the average of the pixel value weighted by the Gaussian function and then multiplied by the exposure time of the respective image. This weighting scheme is designed to give higher weightage to mid-range pixel values and to images with longer exposure times. Oh, Lee, Tai, and Kweon [15] uses rank minimization algorithm using a synthesized multi-exposure LDR image data set used to recover the HDR image. However, this requires the sensor to be linear over its full operating dynamic range to computationally approach an ideal rank-1 structure. Therefore, this method is inherently limiting given its assumption of near ideal linear lower dynamic range sensors. Table 2 data shows that the proposed restricted $v_p$ and exposure time method that does not require any weighting functions for image fusion produces a higher robustness linear HDR image recovery of the test target over the full 78 dB DR versus the tested prior-art methods. Specifically, using the unrestricted $v_p$ prior-art methods, many image sensor pixel $v_p$ readings from the 16 multi-exposure images fall in the non-linear CRF region leading to a non-robust HDR recovery.

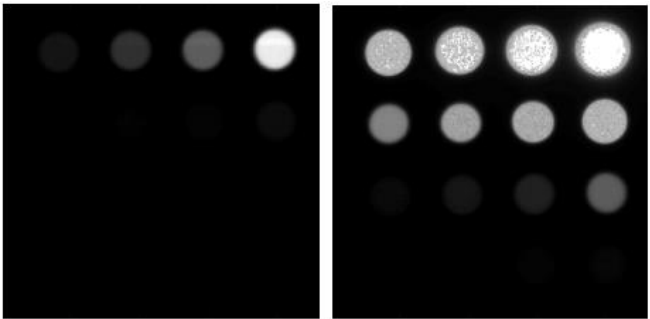

Fig. 8. 78 dB test target images captured for the proposed minimal images multi-exposure method for HDR image recovery. Exposure times are: Left image: 59 μs and Right Image: 5.9 ms.

To test the 80 dB linear DR test target recovery by the proposed and designed minimal 2 images weighting-free multi-exposure method with the experimentally measured CRF of Fig. 6, the 78 dB 16 patch HDR target with 60 KLux LG3 light box illumination was observed using specific exposure times of $T_1$ and $T_2= 100T_1$ that were picked based on the $v_p(avg)$ max/min limitations and desired 80 dB HDR design value given the CRF measured 40 dB $LDR_E$ value. Specifically, a $T_1$=5.9 μs gave a brightest patch $v_p(avg)$ = 28620 which indeed satisfies the linearity limit condition for this camera that the brightest patch $v_p(avg)$ = $v_B$ ≤ $v_{max}$ = 37486. Per design rules, $T_2$ was set to $100T_1$= 5.9 ms and a second scene snap shot image was taken, with both $T_1$ and $T_2$ exposure images shown in Fig. 8. Using individual pixel $v_p$ limits of $v_{max}$=37500 and $v_{min}$=390, the pixel data from the two raw images was filtered to a smaller data set that was linearly transformed to the scaled irradiance values using the measured CRF of Fig. 6. Given a factor of 100 shorter exposure time for the $T_1$ image data, its scaled irradiance values were multiplied by a factor of 100 to put both scaled irradiance data sets from the two captured images on the same relative irradiance scale. Without using any weighting scheme, the two data sets were added to produce the composite linear HDR image. As a 78 dB DR image cannot be displayed, Table 3 shows the recovered HDR image computed spatially averaged (i.e., over a 697 pixels circular zone) patch target DR value readings from the target patch locations. Indeed, the proposed method delivered an accurate and robust recovery of the 16 patch 78 dB target over the full designed 80 dB HDR recovery range.

It is important to test a designed linear HDR camera operation using the proposed scene over a large illumination range to check when the camera system breaks down the input-to-output linear transformation of the proposed minimal images weighting-free multi-exposure HDR recovery technique. In effect the questions being asked are: If the light illumination level decreases a lot, one has to increase the exposure time to a larger value where the sensor may no longer responds linearly, i.e., increase in exposure time does not linearly increase the $v_p$ within the designated 40 dB CRF linear range. Furthermore, if the light illumination level increases a lot, one has to decrease the exposure time to a smaller value where the sensor may also no longer responds linearly, i.e., decrease in exposure time does not linearly decrease the $v_p$ within the designated 40 dB CRF linear range. In addition, the digital sensor shortest exposure time required to meet the brightest patch $v_p(avg) = v_B$ ≤ $v_{max}$ condition may not be possible for the specific sensor. Hence testing is required with different illumination levels to measure the HDR recovery of the 78 dB test target. Table 4 shows the measured results for the current camera system with illumination levels of the LG3 light box changed from 200 KLux to 683 Lux. Data shows a robust HDR recovery for the 68 KLux to 3400 Lux range which is a factor of 20 change in average uniform light illumination. Note that at 200 KLux and with the sensor shortest $T_1$= 29 μs, the acquired short exposure image fails to satisfy the brightest target patch $v_p(avg) = v_B$ ≤ $v_{max}$ = 37486 condition required for the proposed design linear HDR multi-exposure camera system. The LG3 light box has no settings between 200 KLux and 68 KLux, hence no imaging readings are taken within this range to determine exactly where between 200 KLux and 68 KLux does the recovery become non-robust. Note that today, some digital image sensor-based camera systems (e.g., DSLR cameras) are designed with a built-in light meter to measure illumination levels, and the meter's readings can be used to guide the proposed minimal exposures camera HDR-mode operations.

**Table 4. Proposed technique 78 DB target HDR recovery using different illumination levels.**

| Design (dB) | 68 KLux | 60 KLux | 30 KLux | 20.43 KLux | 6800 Lux | 3400 Lux | 1368 Lux | 683 Lux |
|---|---|---|---|---|---|---|---|---|
| 0 | 0 | 0 | 0 | 0 | 0 | 0 | 0 | 0 |
| 8 | 10 | 10.1 | 10 | 10 | 10 | 10.1 | 5.8 | 6.8 |
| 14 | 16.2 | 15.7 | 16 | 16 | 16.1 | 16.3 | 12.9 | 13 |
| 20 | 22.5 | 21.9 | 22.1 | 22.2 | 22.4 | 23.1 | 21.3 | 18.8 |
| 26 | 26.5 | 26.5 | 26.8 | 26.6 | 27.4 | 28.1 | 25.6 | 22.4 |
| 32 | 35.1 | 35.2 | 35.1 | 35.1 | 35.9 | 36.3 | 33.3 | 27.7 |
| 36 | 40 | 40.2 | 40.2 | 39.9 | 39.7 | 40 | 36.7 | 32 |
| 40 | 42.5 | 41.8 | 41.8 | 42.1 | 42.3 | 42.6 | 39.3 | 35 |
| 44 | 44.2 | 43.7 | 43.9 | 43.9 | 43.9 | 44.1 | 41.2 | 37.9 |
| 50 | 53.9 | 53.5 | 53.7 | 53.7 | 53.7 | 53.8 | 49.8 | 45.4 |
| 56 | 56.8 | 56.5 | 56.6 | 56.7 | 56.6 | 56.7 | 52.7 | 48.6 |
| 60 | 61.3 | 61.1 | 61 | 60.9 | 60.9 | 61 | 57.3 | 53.7 |
| 64 | 64.2 | 64.1 | 63.8 | 63.8 | 63.7 | 63.8 | 60.1 | 55.8 |
| 68 | 67.5 | 68 | 67.6 | 67.6 | 67.5 | 67.5 | 62.8 | 59.7 |
| 74 | 73 | 73.4 | 72.6 | 72.5 | 72.4 | 72.8 | 68.8 | 60.9 |
| 78 | 76.1 | 77.4 | 75.9 | 76 | 75.9 | 76 | 71.8 | 59.5 |

Computational HDR imaging methods are widely deployed in photography where one cannot quantitatively verify the natural scene and visual effects are important. Keeping this aspect in mind and for further validation of the proposed method, an additional two experiments are carried out that are more in-line with natural scenes. Given the proposed method is suited for HDR scenes, a hybrid calibrated-natural design scene is created using two DR controlled light patches within a lighted room with a toy car and horse within the field-of-view of the deployed sCMOS camera. The LG3 lightbox at a 60Klux rating using two circular patch zones placed on the lightbox illumination plane create a 0 dB and 78 dB DR rating in the scene. The 0 dB brightest light patch is an open aperture while the 78 dB weakest light patch is made using ND attenuation filters. Using the proposed HDR method, two images of this scene are captured and then processed for HDR image recovery of a 78 dB near natural scene. Fig.9 shows the under-exposed image taken using $T_1 = 5.9$ μs with a $v_p(avg) = 31822$ for the brightest 0 dB patch meeting the $v_p(avg) = v_B \leq v_{max} = 37486$. Fig.9 also shows the over-exposed image taken using $T_2 = 100T_1 = 5.9$ ms with a $v_p(avg) = 800.9$ for the weakest 0 dB patch meeting the $v_p(avg) = v_w \geq v_{min} = 390$. Fig.10 shows the successful recovery of the hybrid design 78 dB DR test scene with a measured bright-to-weak patch ratio of 76.92 dB versus 78 dB ground-truth. The Fig.9 and Fig.10 images are presented in the log scale for ease of viewing of the scene contents and importantly the recovered 78 dB attenuation weak light spot in Fig.10 that appears correctly just above the car roof.

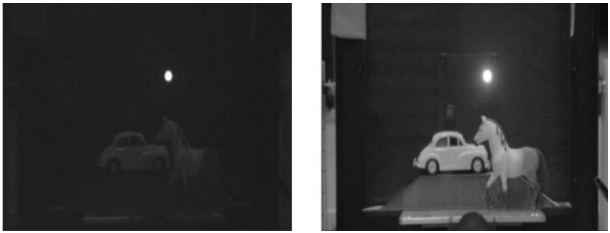

Fig. 9. Hybrid calibrated-natural 78 dB test scene images captured for the proposed minimal images multi-exposure method for HDR image recovery. Exposure times: Left image: 59 μs and Right Image: 5.9 ms.

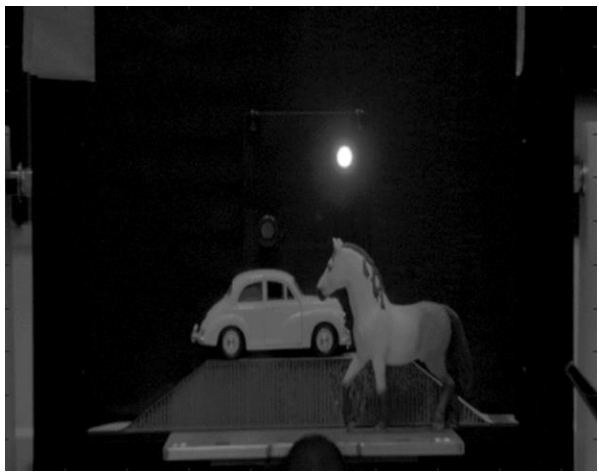

Fig. 10. Recovered hybrid calibrated-natural 78 dB test scene image using the proposed minimal 2 images multi-exposure method showing the 76.9 dB measured attenuation dark spot just above the car roof.

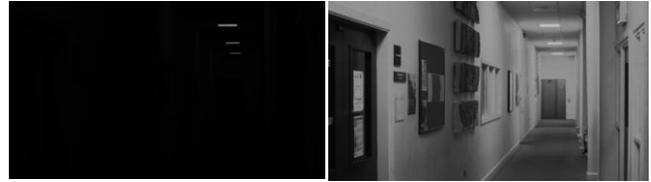

Fig. 11. Indoor natural test scene images captured for the proposed minimal images multi-exposure method for HDR image recovery. Exposure times: Left image: 888 μs and Right Image: 88.8 ms.

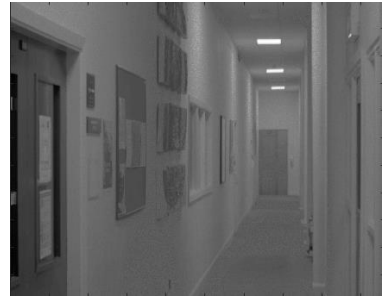

Fig. 12. Recovered natural uncalibrated indoor test scene image using the proposed minimal 2 images multi-exposure method. Fig.11 images-based computation gives an estimated natural scene DR of 49.4 dB.

Next, Fig.11 shows a fully natural but uncalibrated indoor scene test target captured images using $T_1$ and $T_2$ exposure times. $T_2 = 8.8$ ms is chosen first so that the darkest pixels of the scene representing the room door dark region nearest to the camera have a $v_p(avg) = v_w = 1200 \geq v_{min} = 390$. The $T_1$ short exposure per 78 dB calibrated test image camera design is $T_1 = T_2/P_2 = 8.8$ μs as the designed time factor $P_2 = 100$. The brightest pixels of the $T_1$ image using the ceiling lights region gives a $v_p(avg) = 3540 \leq v_{max} = 37486$ and $v_p(avg) = 3540 \geq v_{min} = 390$. As linear CRF camera operation of both images has been maintained using the proposed HDR technique, the bright ceiling lights zone can be estimated to have a $v_p(avg) = 3540 P_2 = 354000$ versus the scene door dark zone $v_p(avg) = 1200$ giving an estimated target scene DR of $20\log(35400/1200) = 49.4$ dB which is not generally considered in the HDR zone although 49.4 dB exceeds the 40 dB linear CRF DR range of the deployed camera. Hence the proposed HDR method still applies for robust linear irradiance range capture of the observed uncalibrated (i.e., unknown ground truth) scene. Fig.12 shows the log scale recovered image of the scene using the two Fig.11 linear 16-bit scale images data processed via the proposed HDR method that can be also be analyzed for visual effects if desired by the photography community.

## 4. Conclusion

Experiments show that the brighter light captured image regions of the tested white light s-CMOS camera shows an individual pixel triggering behavior giving some saturated pixel outputs despite the uniform under saturation light illumination levels. To counter this hardware anomaly that is not a sensor product defect and may be present in other digital image sensors, proposed and demonstrated is a CRF generation technique based on pixel output spatial averaging and histogram analysis for saturated pixel detection, along with the use of an optimized CMOS sensor DR specification limited CRF calibration target suited for low contrast detection applications within a HDR. Specifically, the proposed CRF generation method allows robust estimation of the camera's experimental linear DR region using a highly calibrated "known ground truth" test target with multiple known low contrast zones with a high linear DR. Such a robust calibration method for truest CRF generation is vital for the proposed minimal images multi-exposure weighting-free linear DR extension technique suited for low linear DR digital sensor cameras where linearity must be preserved over the full image signaling chain from image capture to image processing to image display. Experiments have been successfully conducted for both the proposed CRF generation method as well as the linear DR extension image processing technique.

Specifically, experimental CRF data using the proposed CRF measurement scheme shows the deployed 16-bit CMOS sensor to have a highly non-linear response for the brighter regions and a near linear 40 dB DR response between a specific CMOS individual pixel voltage output range from 37486 and 391. The measured CRF is used with 5 leading prior-art multi-exposure HDR image recovery algorithms using 16 exposures for a 78 dB DR test target recovery. In addition, the measured CRF is used with the proposed multi-exposure method using 16 exposures as well as the optimal minimal 2 exposures for allowing a 80 dB HDR recovery. In addition, the proposed minimal exposures unweighted HDR recovery method is successfully tested using a factor of 20 change in the target illumination level. Furthermore, two additional experiments provide validation of the proposed HDR technique by first using a hybrid 78 dB calibrated-indoor natural scene target and next by engaging an uncalibrated indoor natural scene target.

To summarize, the conducted experiments indeed show that the proposed methods for both CRF generation and HDR recovery have higher robustness to non-linearities in the CMOS sensor and deploy the minimal different exposure images and data sets needed to implement multi-exposure image fusion techniques. In addition, the CRF calibration process avoids use of unknown growth truth test image data that introduce uncertainty in the camera imaging operations that can have a detrimental impact for HDR camera measurement science applications. The proposed camera calibration and linear DR extension methods can have impact across numerous applications where limited linearity and DR of digital image sensors hinder the linear HDR imaging capacity of camera systems. Future work relates to testing the proposed methods using a variety of digital sensor camera systems. In addition, future work would involve using higher DR calibrated multiple HDR targets as well as uncalibrated HDR scenes in real indoor and outdoor scenarios.

**Acknowledgment**. The authors thank PhD student Mohsin A. Mazhar for experimental data support.